\documentclass{aa} 
\usepackage{psfig}
\begin{document}

   \thesaurus{02     
              (09.19.1;  
               11.05.2;  
               11.19.3;  
               12.03.3;  
               13.09.3;  
               13.09.1)} 
   \title{ISO deep far-infrared survey in the ``Lockman Hole"}

   \subtitle{II. Power spectrum analysis: evidence of a strong evolution
	in number counts
	\thanks{Based on observations with ISO, an ESA project
	with instruments funded by ESA member states (especially the
	PI countries: France, Germany, the Netherlands, and the United
	Kingdom) and with the participation of ISAS and NASA.}
	\fnmsep\thanks{The ISOPHOT data presented in this paper was 
	reduced using PIA, which is a joint development by ESA 
	Astrophysics Division and the ISOPHOT consortium.}
	}

   \author{H.\,Matsuhara\inst{1} \and K.\,Kawara\inst{2} \and
	Y.\,Sato\inst{1} \and Y.\,Taniguchi\inst{3} \and
	H.\,Okuda\inst{1} \and T.\,Matsumoto\inst{1} \and
	Y.\,Sofue\inst{2} \and K.\,Wakamatsu\inst{4} \and
	L.L.\,Cowie\inst{5} \and R.D.\,Joseph\inst{5} \and
 	D.B.\,Sanders\inst{5}
          }

   \offprints{maruma@astro.isas.ac.jp}

   \institute{The Institute of Space and Astronautical Science(ISAS),
              3-1-1 Yoshinodai, Sagamihara, Kanagawa, 229-8510, Japan
          \and
	      Institute of Astronomy, The University of Tokyo, 
	      2-21-1 Osawa, Mitaka, Tokyo, 181-8588, Japan
	  \and
	      Astronomical Institute, Tohoku University,
	      Aoba, Sendai 980-8578, Japan
	  \and
	      Faculty of Engineering, Gifu University, 
	      Gifu 501-1193, Japan
	  \and
	      Institute for Astronomy, University of Hawaii, 
	      2680 Woodlawn Drive, Honolulu, HI 96822, USA
             }

   \date{Received  9 February 2000 ; accepted 27 June 2000  }

   \maketitle

   \begin{abstract}

We investigate the characteristics of FIR brightness fluctuations at 90\,$\mu$m
and 170\,$\mu$m in the Lockman Hole, which were surveyed with ISOPHOT
aboard the Infrared Space Observatory(ISO).
We first calculated the angular correlation function of each field and then
 its Fourier transform(the angular Power 
Spectral Density: PSD) over the spatial frequency range of $f=0.05 - 1
$~arcmin$^{-1}$.
The PSDs are found to be rather flat at low spatial frequencies($f \leq
0.1 \, \mathrm{arcmin^{-1}}$), slowly decreasing toward higher 
frequencies.
These spectra are unlike the power-law ones seen in the IR cirrus
fluctuations, and are well explained by randomly distributed 
point sources.  Furthermore, point-to-point comparison
between 90\,$\mu$m and 170\,$\mu$m brightness shows a linear
correlation between them, and the slope of the linear
fit is much shallower than that expected from the IR cirrus color, and is 
consistent with the 
color of galaxies at low or moderate redshift($z<1$). We conclude that the
brightness fluctuations in the Lockman Hole are not caused by the IR cirrus, but are 
most likely due to faint star-forming galaxies. We also give the constraints on the 
galaxy number counts down to 35 mJy at 90\,$\mu$m and 60 mJy
at 170\,$\mu$m, which 
indicate the existence of a strong evolution down to these fluxes in the
counts. The galaxies responsible for the fluctuations also significantly
contribute to the cosmic infrared background radiation.  
      \keywords{Galaxies: evolution --
                Galaxies: starburst --
		Cosmology: observations --
		Infrared: ISM: continuum --
                Infrared: galaxies
               }
   \end{abstract}

%
\section{Introduction}

In order to understand the history of galaxy formation and evolution, surveys 
at far-infrared(FIR) and submillimeter wavelengths are essentially important, since a 
large portion of star formation activity in the universe
may be hidden by  dust, prohibiting optical and near-infrared studies due 
to the enormous extinction. The recent detection of the far-infrared Cosmic Infrared Background(CIB)
radiation by COBE, which appears to have comparable brightness 
to the total intensity in deep optical counts from the Hubble deep field
(Fixen et~al.  \cite{Fix98}; Hauser et~al. \cite{Hau98}; Lagache et~al. \cite{Lag00b}), 
indicates that the infrared-bright galaxies are responsible 
for roughly half of the energy released by nucleosynthesis. 
Various discrete source surveys are also being pursued from mid-infrared to 
submillimeter wavelengths. For example, source counts at 15\,$\mu$m 
(Aussel et~al. \cite{Aus98}; Altieri et~al. \cite{Alt98}; Elbaz et~al. \cite{Elb99}),
at 170\,$\mu$m (Kawara et~al. \cite{Kaw98}(paper I) ; Puget et~al. \cite{Pug99}),
and at 850\,$\mu$m (Blain et~al. \cite{Bla99}) have been reported.
Using both CIB and these new number count data, modeling of the cosmic star
formation history has been attempted by many authors(Guiderdoni et~al. \cite{Gui98};
Dwek et~al. \cite{Dwe98}; Rowan-Robinson \cite{RR99}; Tan et~al. \cite{Tan99};
Ishii et~al. \cite{IT99}). All of them require a strong evolution in the 
star formation rate as we look back to high redshift(more than 10 times larger
at $z \simeq 1$).

Although the FIR deep survey is now considered a key observing method 
for the exploration of the ``optically dark side" of the star formation history
of galaxies(Guiderdoni 
et~al. \cite{Gui97}), the detectivity of 1m-class space FIR telescopes is likely to 
be limited by the noise due to the fluctuation of the IR cirrus(Low 
et~al. \cite{Low84}), emission from the interstellar dust,  even at high Galactic 
latitude(Helou \& Beichman
\cite{HeBe90}). Hence a detailed study of the IR cirrus fluctuations is highly
important, especially for the planning of the deep surveys 
intended with forthcoming IR space
telescopes such as IRIS(Murakami \cite{Mur98}) and SIRTF(Fanson et al.
\cite{Fan98}).

The spatial structure of the IR cirrus at 100\,$\mu$m as
measured by IRAS was extensively studied by Gautier et~al. (\cite{Gau92})
and Abergel et~al.(\cite{Abe96}). 
From a Fourier analysis of the brightness distribution, Gautier et~al. found 
that the Power Spectral 
Density(PSD) of the brightness fluctuation at 100\,$\mu$m follows a 
power-law function of the spatial frequency with an index of about
3 below the spatial frequency corresponding to the IRAS
beam size(about $0.25\, \mathrm{arcmin}^{-1}$). They also found that the PSD
is proportional to $B_{\rm 0}^{\rm 3}$, where $B_{\rm 0}$ is 
the mean brightness of the IR cirrus.
The ISOPHOT(Lemke et~al. \cite{Lem96}) onboard the ISO(Kessler et~al. 
\cite{Kes96}) is capable of observing the 
IR cirrus at wavelengths longer 
than 100 $\mu$m, with a better spatial resolution than that of IRAS.
Herbstmeier et~al. (\cite{Her98}) analyzed the spatial 
characteristics of four fields measured by ISOPHOT, and obtained similar 
power-law spectra for relatively bright cirrus regions.
Recently Lagache \& Puget (\cite{Lag00a}) have made a power spectrum
analysis of the 170\,$\mu$m image of the Marano 1 field, and
found a significant excess at $f=0.25 - 0.6\, \mathrm{arcmin}^{-1}$, 
which they attributed to
the fluctuations due to unresolved extra-galactic sources.

In this paper we investigate the characteristics of FIR brightness
fluctuations in the Lockman Hole, a region with a uniquely low HI 
column density(Lockman et~al. \cite{Loc86}), and thus minimal IR cirrus 
contribution to the fluctuations. Therefore the fluctuations in the Lockman
Hole are likely to be dominated by faint, distant galaxies(Helou \& Beichman 
\cite{HeBe90}; Herbstmeier et~al. \cite{Her98}) and the fluctuation
analysis of FIR images will provide unique
information on the number counts of infrared galaxies even below 
the source-confusion limit, thus constraining the parameters
characterizing the number count models.

This paper is organized as follows: section 2 briefly describes the
observations and data processing. Section 3 explains the power spectrum analysis
and examines the contribution by the IR cirrus fluctuations and also
describes a simulation of the images and the PSDs.
Section 4 describes the constraints on the 
galaxy number counts. The nature of the sources responsible for
the fluctuations is
discussed in section 5, and section 6 gives the conclusions.

\begin{figure*}
\vbox{\vspace{0cm}\psfig{figure=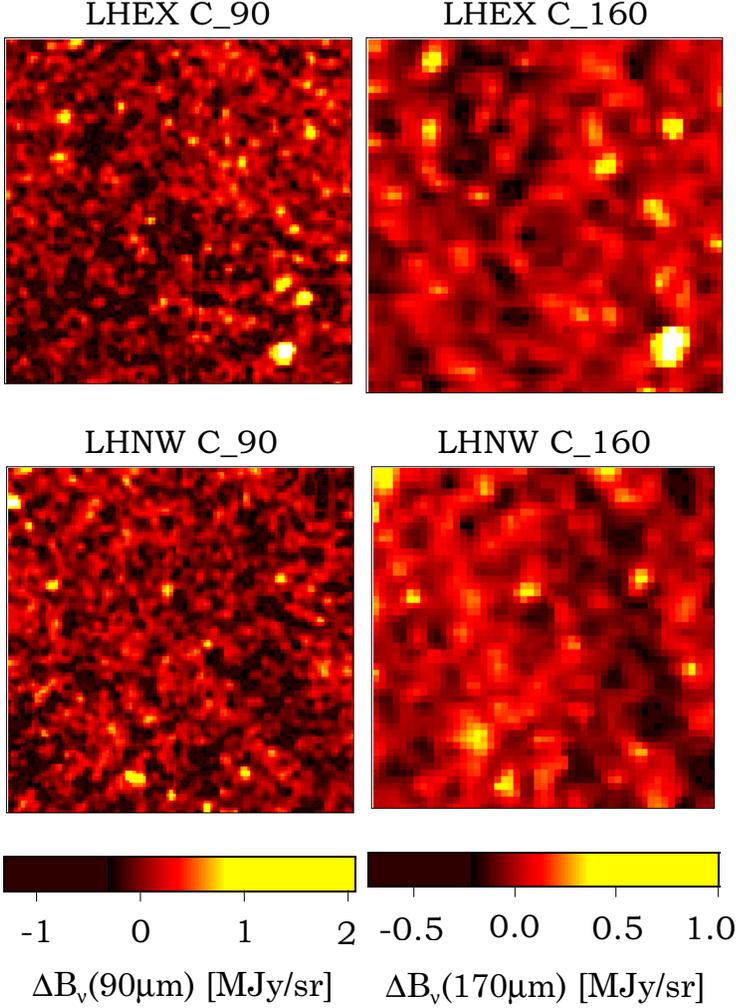,width=11.0cm,clip=yes}\vspace{-5cm}}
\hfill\parbox[b]{6.5cm}{\caption[]{The left column shows 90$\mu$m
	images of LHEX(top) and LHNW(bottom). Each image is $39\arcmin.5
	\times 39\arcmin.5$ wide($\rm 23\arcsec / pixel$). The right 
	column shows 170$\mu$m images($\rm  46\arcsec / pixel$). The LHEX
	 image is $40\arcmin.6 \times 40\arcmin.6$ 
	wide, while the LHNW one is $39\arcmin.1 \times 39\arcmin.1$ wide. 
	The brightness shown in each image is offset 
	from its median brightness.
	Each image is made of four $22\arcmin \times 22\arcmin$ sub-fields.
	Roughly the north is left and the west is top.   }}
 \label{fig:fig1}
\end{figure*} 

\section{Observation and data processing}

The FIR survey of the Lockman Hole, which was executed
as a part of Japan/UH cosmology program using the ISAS guaranteed time,
is described in Taniguchi et~al. (\cite{Tan94}) and paper I. 
Two $44\arcmin \times 44\arcmin$ fields named LHEX and LHNW were
mapped with two filters: C\_90(centered at 90\,$\mu$m) and 
C\_160(170\,$\mu$m). Each of the two fields is made up of 4 sub-fields. 

Each sub-field map was produced from the edited raw data by 
the PHT Interactive Analysis(PIA; Gabriel et~al. \cite{Gab97}) 
version 7.1 or 7.2, and is  hereafter referred to as an 
AAP(Astronomical Analysis Processing of PIA) map. Each AAP map
is either $58 \times 58$ pixels($\rm 23\arcsec/pixel$) for a 90\,$\mu$m 
sub-field, or $31 \times 31$ pixels($\rm 46\arcsec/pixel$) for a 
170\,$\mu$m sub-field.
To correct the drift in the responsivity of the detectors, we applied 
the median filter smoothing (see Paper I for details) to the AAP data.

Together with the final AAP maps we also produced the uncertainty maps,
and found that the typical 3$\sigma$ noise is as low as 0.012MJy/sr( 0.60mJy/pixel 
for a $46\arcsec \times 46\arcsec$ pixel), indicating that the instrumental
noise is negligible in the following results.

As was done in Paper I,  the observed fluxes as well as the brightness 
of the images are scaled based on the fluxes
of the brightest source \object{F10507+5723}(\object{UGC 06009}) measured 
with IRAS. We found that the flux calibration based on the FCS1 
measurements underestimates the 90\,$\mu$m flux of the IRAS source 
by a factor of 2.6, 
although it gives the mean brightness of the images consistent with the 
COBE/DIRBE brightness within 25 per~cent. The 170\,$\mu$m
flux of the IRAS source is assumed to be 1133 mJy as described in Paper I.
The flux calibration based on the FCS1 measurements again underestimates
the 170\,$\mu$m flux by a factor of 1.5. For the 170\,$\mu$m flux, we
found that the discrepancy is mostly
due to an underestimate of the effective solid angle of the ISOPHOT 
detector. Puget et~al. (\cite{Pug99}) and Lagache \& Puget (\cite{Lag00a})
used an effective solid angle at 170\,$\mu$m derived from Saturn footprint
measurements, which is significantly larger than that used in the PIA.
Here we assume theoretical PSFs of a telescope with a 60 cm 
primary mirror and a 20 cm secondary mirror which agree well with those
measured during the ISOPHOT calibration observations(M\"uller~\cite{Mul99}).
With this assumption, the factors of 2.6(90\,$\mu$m) and 1.5(170\,$\mu$m) 
discrepancies reduce
to factors of 2.0(90\,$\mu$m) and 1.2(170\,$\mu$m). The origin of the 
residual discrepancies is still unknown. However, this is not
problematic for the main results and conclusions of this paper as 
described in section 4 and 5 as long as the fluxes of \object{F10507+5723} 
determined by the IRAS faint source survey are correct.
\begin{figure}
\vbox{\vspace{0cm}\psfig{figure=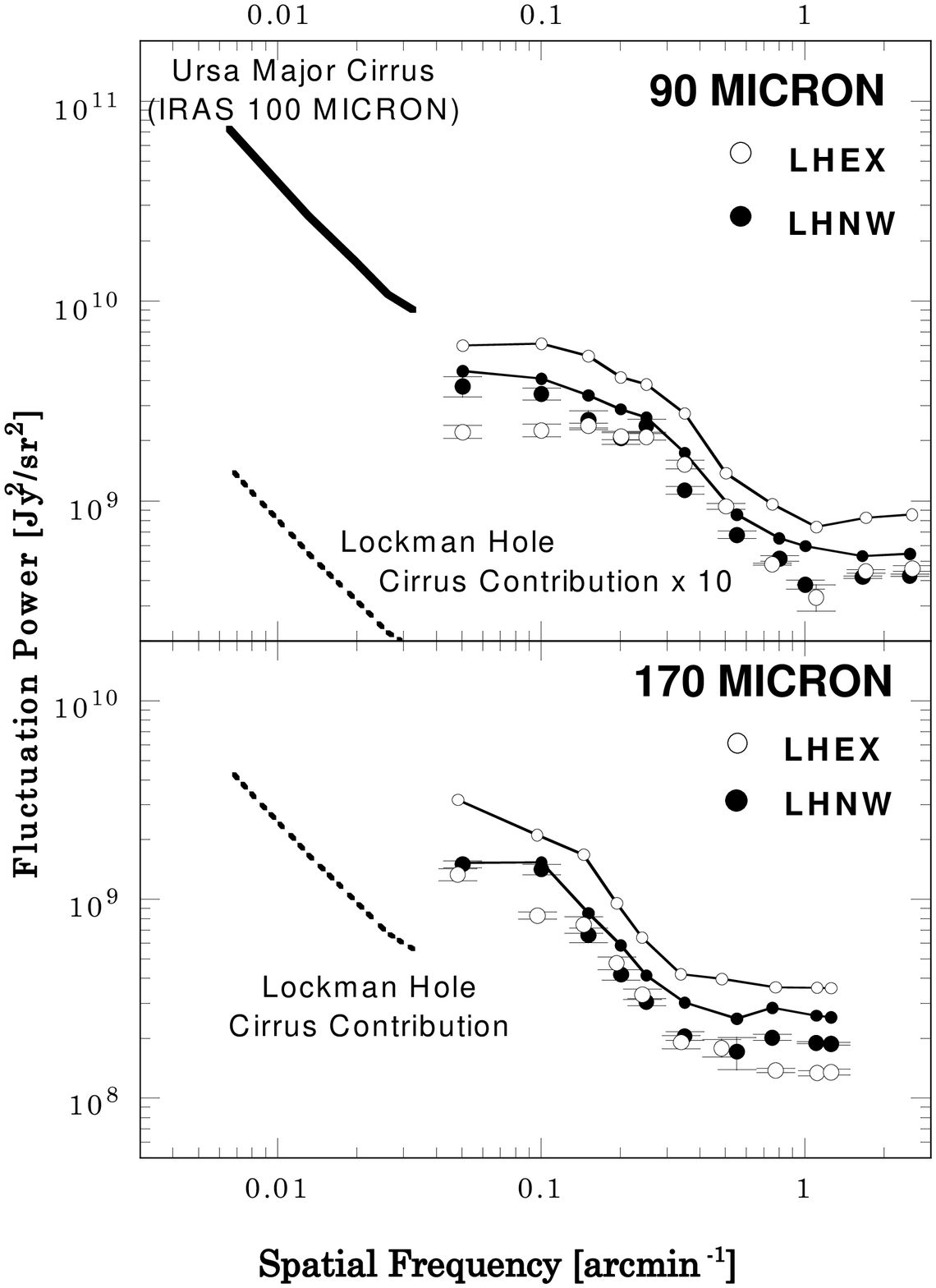,width=8.8cm,clip=}\vspace{0cm}}
\parbox[b]{8.8cm}{\caption[]{Fluctuation power spectral densities(PSDs)
of 90\,$\mu$m(top) and 170\,$\mu$m(bottom) images. Open circles 
represent PSDs of LHEX and filled circles represent PSDs of LHNW. 
As well as the PSDs of the original images(Fig.~\ref{fig:fig1}) shown by circles
connected by thin-solid lines, the PSDs of the residual images
(``residual PSDs"), where
the pixels containing bright sources above 150 mJy(90\,$\mu$m), 250 mJy
(170\,$\mu$m) are masked, are also shown by
circles alone. The brightest source (\object{IRAS F10597+5723})
located in LHEX significantly contributes to the PSDs of the LHEX
images. The 1$\sigma$ error bars, shown only for the residual PSDs, 
represent the standard deviation in a set of the PSDs with different
position angles in the sky.  Thick line is an example of the IR cirrus PSD, 
which is an average spectrum of several IR cirrus in Ursa Major, with 
100\,$\mu$m brightness of 2--3 MJy/sr. Dotted lines are the IR cirrus PSDs 
in the Lockman Hole, estimated
by assuming that the cirrus PSDs are proportional to $B_{0}^{3}$, where
$B_0$ is mean brightness of the cirrus cloud. }}
\label{fpsdcir}
\end{figure} 

Fig.~\ref{fig:fig1} shows the images of LHEX and LHNW used for 
the fluctuation analysis, each of which is the largest square area extracted 
from the mosaiced map(see Fig. 3 of Paper I) made up from four sub-field 
AAP maps. Each image is rebinned into $103 \times 103$~pixels$^2$
(C\_90), $53 \times 53$~pixels$^2$(C\_160 LHEX), or $51 \times 
51$~pixels$^2$(C\_160 LHNW). The plate scale is $23\arcsec/\mathrm{pixel}$ 
for 90\,$\mu$m maps, and $46\arcsec/\mathrm{pixel}$ for 170\,$\mu$m maps. 

\section{The power spectra and the simulation}

For each image shown in 
Fig.~\ref{fig:fig1} the 2-dimensional angular
correlation function $C(x,y)$ was calculated from the brightness 
distribution $B(x_0, y_0)$:
\begin{equation}
C(x,y)= <(B(x_0, y_0) - \bar{B})(B(x_0 + x, y_0 + y) - \bar{B})>
\label{eq:equation1}
\end{equation}
where bracket $< \, >$ represents the average over the whole area in each
image shown in Fig.~\ref{fig:fig1}, and $\bar{B} = <B>$. 
Then, the coordinates were expressed in polar coordinates as  $C(r, \theta)$,
and the Fourier transforms in various radial directions  were calculated
({\it i.e.} $\theta$ is fixed for each transform).
One-dimensional power spectral density(PSD) $P(f)$ is calculated from
\begin{equation}
  P(f) = {2 \over N} \sum_{k=0}^{N-1} C_k cos({2 \pi f r_{k}}) \, ,
\end{equation}
where $f$  is spatial frequency($ 1/(Nr_{\mathrm{pix}}) \leq f \leq 1/(2r_{\mathrm{pix}})$),
$r_{k} \equiv r_{\mathrm{pix}}k$ ($k=0, 1, \cdots, N-1$), 
$r_{\mathrm{pix}}$ is the pixel size of the map, $Nr_{\mathrm{pix}}$ is the largest 
angular size for which the angular correlation function is evaluated, and
$C_k \equiv C(r=r_{k}, \> \theta)$ . Note that 
the units of the PSDs are the same as that of the angular correlation function:
$\rm Jy^{2} /sr^{2}$.
Finally, the PSDs were averaged with respect to $\theta$ 
 and are shown in Fig.~\ref{fpsdcir}. It is noteworthy that the
fluctuations at high frequencies($f \ge 
0.8$~arcmin$^{-1}$ for 90\,$\mu$m, and $f \ge 0.4$~arcmin$^{-1}$ for 
170\,$\mu$m) are smoothed by the instrumental beam and therefore
the PSDs decrease appreciably.
The small error bars of PSDs represent the standard deviation 
among a set of the PSDs with different $\theta$, showing that the PSDs
are almost independent of $\theta$.

In order to check the contributions from bright sources,
the PSDs are derived by masking circular regions with a 4$\times$FWHM
diameter around bright sources with fluxes above $S_{\mathrm{max}}$:
$S_{\mathrm{max}}=250$ mJy at 170\,$\mu$m\footnote{ As for 170\,$\mu$m
images, smaller values of $S_{\mathrm{max}}$ significantly reduce the
amount of residual images.},
$S_{\mathrm{max}}=150$ mJy at 90\,$\mu$m.
In the following the resultant PSDs are called  
``residual PSDs", and are also shown in Fig.~\ref{fpsdcir}.
Interestingly, the residual PSD for each image is more than half of 
the PSD of corresponding original image with almost the same spectral shape, 
indicating that there remains significant contribution from randomly 
distributed point sources with fluxes below $S_{\mathrm{max}}$.

In Fig.~\ref{fpsdcir} typical IR cirrus PSDs are also compared in
order to check the contribution of the IR cirrus to the PSDs of the 
Lockman Hole\footnote{The formula for the IR cirrus PSD by Gautier et~al.
(\cite{Gau92}) cannot be directly compared with the present work, because
the power spectrum analysis presented here is one-dimensional one while
the formula of Gautier et~al. is based on the two-dimensional analysis.}. 
We examined
several IRAS 100\,$\mu$m maps of high-latitude clouds in Ursa Major($l=
 145^{\circ}$, $b=40^{\circ}$), which are reproduced from the IRAS
Sky Survey Atlas(ISSA) by reducing the brightness by a factor of 
0.72, following the COBE/DIRBE calibration(Wheelock et~al. \cite{Whe94}).
The average brightness of the cirrus is 2--3 MJy/sr, and each map
is $150\arcmin \times 150\arcmin$ wide with $1\arcmin.5/\mathrm{pixel}$. The 
derived PSDs show a power-law distribution with an index of -1.5.
Gautier et~al.(\cite{Gau92}) noted that one-dimensional analysis of the IR 
cirrus yielded spectral indices near -2.  These cirrus fluctuation spectra
are much different from those obtained for the Lockman Hole 
images at 90\,$\mu$m, which
show rather flat spectra at lower spatial frequencies. The 170\,$\mu$m
spectra present a slope similar to the IR cirrus one, but this can be
explained by the shape of the footprint power spectrum of ISOPHOT detectors. 
Moreover, the fluctuation powers are much larger than those estimated for the 
IR cirrus, which are also shown in Fig.~\ref{fpsdcir}.
We assume that the cirrus PSD is proportional 
to $B_{\rm 0}^{\rm 3}$(Gautier et~al. \cite{Gau92}), and taking the mean
brightness $B_{\rm 0}$ of the IR cirrus in the Lockman Hole as
0.33 MJy/sr at 90\,$\mu$m and 1.0 MJy/sr at 170\,$\mu$m. These values are
estimated from the atomic hydrogen column density of $6 \times 10^{19} \,
\mathrm{atoms \> cm^{-2}}$ in LHEX(Jahoda et~al. \cite{Jah90})
and the COBE/DIRBE data analysis in the Lockman Hole by Lagache et~al. 
(\cite{Lag99a}, in their Table~4).
We can conclude that the IR 
cirrus contribution to the PSDs in the Lockman Hole is negligible
over all spatial frequencies $f \geq 0.05$.

\begin{figure}
\vbox{\vspace{0cm}\psfig{figure=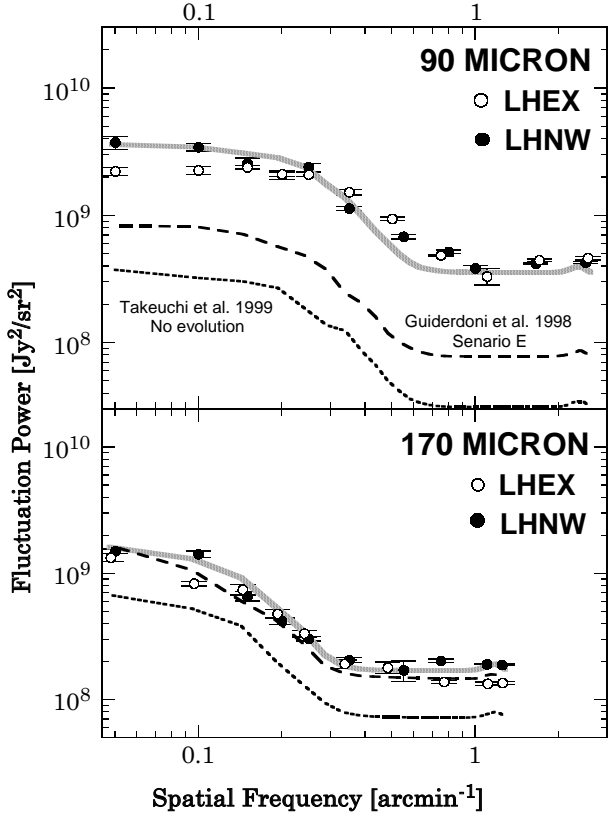,width=8.8cm,clip=}\vspace{0cm}}
\parbox[b]{8.8cm}{\caption[]{The residual PSDs of 
 90\,$\mu$m(top) and 170\,$\mu$m(bottom) images(open circles 
and filled circles, same as Fig.~\ref{fpsdcir}) are compared 
with the simulated PSDs based on various number counts models:
 dashed lines are PSDs of the simulated images by Guiderdoni et~al.
(\cite{Gui98}) scenario E, while the dotted lines are those by Takeuchi
et~al.(\cite{Tak99}) no-evolution. Thick gray lines are examples of
the simulated images produced by simple double power-law number count
models(see text for details). }}
 \label{fpsdgal}
\end{figure} 

In the following we interpret the residual PSDs in terms of
unresolved point sources, probably galaxies. We neglect the
spatial correlation between galaxies, thus assuming that galaxies 
are randomly distributed in the images. Then the
residual PSD will be the product of the footprint power spectrum of the 
ISOPHOT detector and the fluctuation power due to the point
sources, which is a white power spectrum given by:
\begin{equation}
  P_{\mathrm{source}}=\int^{S_{\mathrm{max}}}_{S_{\mathrm{min}}} S^{2}{dN \over dS}dS
\label{eq:eq3}
\end{equation}
where $dN/dS$ is differential source counts. From the residual PSDs
observed, we derive $ P_{\mathrm{source}}=13000\pm3000\, \mathrm{Jy^{2}/sr}$ at
90\,$\mu$m ($S_{\mathrm{max}}=150\, \mathrm{mJy}$) and
$ P_{\mathrm{source}}=12000\pm2000\, \mathrm{Jy^{2}/sr}$ at 170\,$\mu$m
($S_{\mathrm{max}}=250 \,\mathrm{mJy}$), where the errors do not
include systematic ones due to uncertainties in the flux calibration.
 Lagache \& Puget (\cite{Lag00a})
reported the detection of $P_{\mathrm{source}}=7400\, \mathrm{Jy^{2}/sr}$
at 170\,$\mu$m for the Marano 1 field, after sources brighter than 100 mJy
are removed. Contribution from detected sources with fluxes
between 100 mJy and 250 mJy is estimated to be about 7000 $\mathrm{Jy^{2}/sr}$.
Thus the fluctuation power in the Lockman Hole due to the sources 
fainter than 100 mJy is $\sim 5000\, \mathrm{Jy^{2}/sr}$,
 which is comparable to that observed in the Marano 1 field.

 A simulation was performed by making
90\,$\mu$m and 170\,$\mu$m images made up only by galaxies with fluxes
between $S_{\mathrm{min}}$ and $S_{\mathrm{max}}$ and calculating
their PSDs.  Here galaxies are treated as point sources with a PSF
specific to the respective wavelength band of ISOPHOT. 
We used the image of the bright IRAS source (\object{F10507+5723}) seen in
LHEX images as the PSF. The number of sources and their flux densities 
are controlled by the source counts.
We examined the non-evolution count model by Takeuchi et~al.
(\cite{Tak99})  and the scenario E by Guiderdoni et~al.(\cite{Gui98}).
 We assume $S_{\mathrm{min}}=10\,\mathrm{mJy}$ because
the fluctuations due to galaxies fainter than 10 mJy are negligible
in case of these models. The resulted PSDs are compared with the residual
PSDs in Fig.~\ref{fpsdgal}. These simulated
PSDs are not sufficient to explain the observed PSDs although the spectral 
shapes are quite similar. 

\section{Constraints on the Number Counts}

In this section we will investigate the source 
counts below $S_{\mathrm{max}}$, by fitting the simulated PSDs to the 
residual PSDs.
In Paper I we presented the number density of
sources brighter than 150 mJy at both filter bands. 
After Paper I was published, the data processing and the source
extraction technique have been improved, and now we obtained  
source number counts down to $S=70$ mJy at 90\,$\mu$m and $S=100$ mJy at 
170\,$\mu$m, although the counts at such low fluxes may be overestimated
due to the source confusion. We found that the cumulative counts above 
150 mJy show quite a steep increase as the source flux 
decreases: $\beta \simeq 3$ where $N(>S) \propto S^{-\beta}$. These
results will be described in detail in Kawara et~al. (\cite{Kaw00}, 
paper III). If $\beta$ remains equal to -3 down to $S_{\mathrm{min}}
=1\>\mathrm{mJy}$, then the fluctuation power($P_{\mathrm{source}}$) 
calculated by equation~\ref{eq:eq3} 
exceeds the observed ones by an order of magnitude. Hence at a certain
flux below $S_{\mathrm{max}}$, we expect that the slope of the counts
must flatten to $\beta < 2$ so that the predicted fluctuations do not exceed 
the observed ones.    We thus considered simple double power-law count
models and evaluated the
simulated PSDs: in the
flux range $S_{\mathrm{max}} \geq S \geq S_{\mathrm{crit}}$
  \begin{equation}
      N(>S)  =  N_{\mathrm{max}} \left(\frac{S}{S_{\mathrm{max}}}
\right)^{-\beta_{0}} \, , \,
    \label{eq:equation2}
  \end{equation} 
and in flux range $S_{\mathrm{crit}} \geq S \geq S_{\mathrm{min}}$
  \begin{equation}
      N(>S) =   N_{\mathrm{crit}} \left(\frac{S}{S_{\mathrm{crit}}}
\right)^{-\beta_{1}}  \, , \,
    \label{eq:equation3}
  \end{equation}
where $N_{\mathrm{crit}} = N_{\mathrm{max}}(S_{\mathrm{crit}}/S_{\mathrm{max}}
)^{-\beta_{0}}$, and $\beta_{0}$, $\beta_{1}$ are assumed to obey
inequalities $ 0 \leq \beta_{1} < \beta_{0} \leq 4$.
 $N_{\mathrm{max}}$ and its uncertainty are listed in 
Table~\ref{tab:ncp}. The uncertainty in $N_{\mathrm{max}}$ includes
Poisson uncertainties based on the total number of sources with flux
above $S_{\mathrm{max}}$ in all images, and systematic ones originating
from the incompleteness due to the source confusion, which will be
discussed in paper III. 
If $\beta_{1} < 2 < \beta_{0}$, the simulated PSD is not sensitive 
to $S_{\mathrm{min}}$ and is
dominated by sources with fluxes around $S_{\mathrm{crit}}$, as discussed in 
Lagache and Puget (\cite{Lag00b}).
Various sets of parameters can be chosen so that the simulated PSDs
fit to the residual PSDs. 
Examples of the simulated PSDs which fit well to the residual
PSDs are also shown in Fig.~\ref{fpsdgal},
for which $\beta_{0} = 3.0$, $\beta_{1} = 1.0$ for both 90\,$\mu$m 
and 170\,$\mu$m counts, and  $ S_{\mathrm{crit}} = 80 $ mJy
for 90\,$\mu$m counts and $ S_{\mathrm{crit}} = 150 $ mJy for 170\,$\mu$m one.
We found that the simulated PSDs can fit the residual PSDs over all
spatial frequencies within $\pm 22$ per~cent for the 90\,$\mu$m PSDs and $\pm 17$ 
per~cent for the 170\,$\mu$m ones. These deviations of the residual 
PSDs from the simulated PSDs are also included as uncertainties in the
residual PSDs.

\begin{figure}
\vbox{\vspace{0.0cm}\psfig{figure=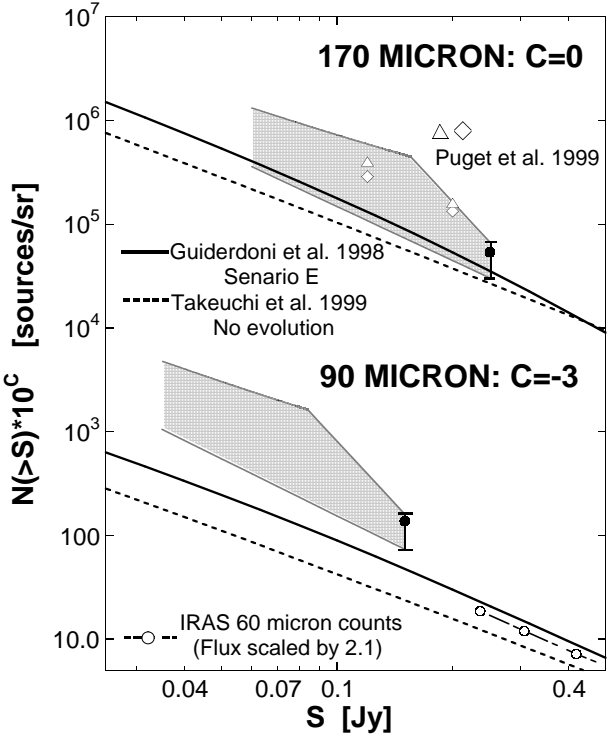,width=8.8cm,clip=}\vspace{0.0cm}}
\parbox[b]{8.8cm}{\caption[]{The shaded areas show the allowed regions
for various number count models, which are consistent
with the fluctuation powers of the Lockman Hole images(Fig.~\ref{fpsdgal}).
The bottom border is obtained from the simple number count
model with a single power-law index between  
$S_{\mathrm{min}}=1$ mJy and $S_{\mathrm{max}}=150$ mJy(90\,$\mu$m) or
250 mJy(170\,$\mu$m). The upper border is determined from the allowed double
power-law count models.
See text for detailed information. The filled
circles are observed number counts at $S_{\mathrm{max}}$. The theoretical 
number count models by Guiderdoni et~al.(\cite{Gui98}) (scenario E : 
solid lines) and the no evolution model by Takeuchi et~al.(\cite{Tak99}) 
(dashed lines) are also shown. The open diamonds and triangles are observed 
and incompleteness-corrected
source counts obtained by the FIRBACK Marano 1 survey(Puget et~al.
\cite{Pug99}). The open circles with dash-dotted line show the IRAS 
60\,$\mu$m counts, in which the flux is scaled by
the factor $S_{\nu}(90\,\mu \mathrm{m})/ S_{\nu}(60\,\mu \mathrm{m}) =
 2.1 $. }}
 \label{fig:gcounts}
\end{figure} 

We now derive the allowed regions in plots of the 
cumulative counts  at fluxes below $S_{\mathrm{max}}$. 
The bottom border of the allowed regions is 
derived by examining the single power-law counts($S_{\mathrm{crit}} =S_{\mathrm{min}}
= 1\,\mathrm{mJy}$)\footnote{Lower value of $S_{\mathrm{min}}$ does not appreciably 
change the results: $\beta_{0}$=1.7 even if $S_{\mathrm{min}} = 0$.}
and we obtain $\beta_{0}$=1.8 for 90\,$\mu$m counts and $\beta_{0}$=1.7
for 170\,$\mu$m ones. In case of double power-law counts we may choose a larger 
value of $\beta_{0}$ with a set of parameter
pairs($S_{\mathrm{crit}}$, $\beta_1$), and among them $S_{\mathrm{crit}}$ for
$\beta_{1} = 0$ is the largest. Thus we determine the upper borders of the allowed 
regions by connecting ($S_{\mathrm{crit}}$, $N_{\mathrm{crit}}$) points of 
numerous models with $\beta_{1} = 0$ and $\beta_{0} = 2 - 4$.
The results are shown by the shaded area in Fig.~\ref{fig:gcounts}.
 As for the 170\,$\mu$m counts, the bottom border of the allowed
region is close to the Scenario E model by Guiderdoni
et~al. (\cite{Gui98}), and is consistent with the counts at 120 mJy
and at 200 mJy obtained from the FIRBACK Marano 1 survey (Puget et~al.
\cite{Pug99}). On the other hand, the allowed region for 90\,$\mu$m 
counts are significantly above those of any currently existing models. 

The bottom border derived here cannot be applied below
a certain flux $S_{\mathrm{L}}$, which is the flux at the intersection
between an upper-border count model with the steepest slope
($\beta_{\mathrm{0}}=4$, $\beta_{\mathrm{1}}=0$), and
the bottom border: $ S_{\mathrm{L}}=30\, 
mJy $(90\,$\mu$m) or 50 mJy(175\,$\mu$m).
By definition, the upper border model do not require sources below
$S_{\mathrm{crit}}$($>S_{\mathrm{L}}$).
On the other hand, in case of the bottom border model,  contribution to the
fluctuations from sources below $S_{\mathrm{L}}$ is still appreciable, about
half of the total. Considering the uncertainties in $N_{\mathrm{max}}$
and the residual PSDs as discussed above, we finally give 
$ S_{\mathrm{L}}=35\,mJy $ for 90\,$\mu$m counts and $ S_{\mathrm{L}}=
60\,mJy $ for 170\,$\mu$m counts.

\begin{table}
   \caption[]{Number count parameters}
	\label{tab:ncp}
   \[
	\begin{array}{lllcl}
	  \hline
	  \noalign{\smallskip}
	    \mathrm{band}  & S_{\mathrm{max}} & N_{\mathrm{max}} & \mathrm{uncertainty}
		  \\
	     &  [\mathrm{Jy}]  &  [\mathrm{sources/sr}]  & \mathrm{in}\, N_{\mathrm{max}} \\
	  \noalign{\smallskip}
	  \hline
	  \noalign{\smallskip}
	    90\, \mu\mathrm{m} & 0.15  &  1.4 \times 10^{5} & +16 \%, -47 \%  \\
	   170\, \mu\mathrm{m} & 0.25  &  5.4 \times 10^{4} & +24 \%, -44 \%  \\
	  \noalign{\smallskip}
	  \hline
	\end{array}
   \]
\end{table}

\section{Discussion}

\begin{figure}
\vbox{\vspace{0.0cm}\psfig{figure=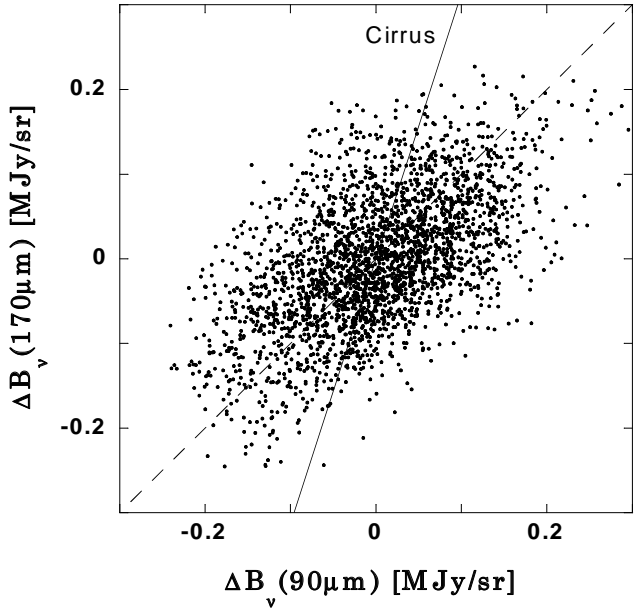,width=8.8cm,clip=}\vspace{0.0cm}}
\parbox[b]{8.8cm}{\caption[]{Point-to-point comparison between the 
90\,$\mu$m sky brightness and the 170\,$\mu$m one in the Lockman
Hole. The brightness is not absolute one, but is offset from the median 
brightness.  In this plot the pixels around the 
bright sources($S \geq 150$mJy for 90\,$\mu$m, $S \geq 250$mJy for 
170\,$\mu$m) are not shown. In order to avoid the error due to the 
difference of the beam size between the two bands, the 90\,$\mu$m  
images are convoluted with the 170\,$\mu$m beam
profile(144\arcsec FWHM).  The solid line showing 
the IR cirrus color vector is taken from Lagache et~al. (\cite{Lag99a}), which is 
17.5K gray-body with an emissivity proportional
to $\lambda^{-2}$.  }}
 \label{fig:f_bvsb}
\end{figure}

We attribute the origin of the residual PSDs to the random distributions of
faint sources which can no longer be identified as individual sources due
to the source confusion.
Another piece of evidence which supports this interpretation
is the 170\,$\mu$m/90\,$\mu$m brightness ratio of the FIR background 
emission. In Fig.~\ref{fig:f_bvsb} we show the point-to-point comparison between
the 90\,$\mu$m and the 170\,$\mu$m sky brightness. Here we examine the
residual images after masking pixels around the sources with $S > 
S_{\mathrm{max}} $.  In order to match the spatial
resolution between 90\,$\mu$m($76\arcsec$ FWHM) and 170\,$\mu$m($144\arcsec$ FWHM)
images, one wavelength band image is convoluted with the other wavelength band beam
profile.  Although the scatter is large, the plot shows linear 
correlation with a slope of unity. This background color
is quite different from that expected for the IR cirrus(about 3.1), and can be
interpreted as a typical FIR color of galaxies contributing to the fluctuation.
The fluctuation color due to faint galaxies depends on their redshift, and
 their SEDs. The relations between the FIR flux ratio and the redshift 
for SEDs of the cirrus dominated galaxy, the pure starburst galaxy(Efstathiou et~al.
\cite{Efs00}; Efstathiou \& Siebenmorgen \cite{ES00}), and the mixture of 
these two SEDs which represents a star-forming galaxy's SED like 
IRAS \object{F10507+5723}, are shown in Fig.~6. The flux ratio of 
$S_{\nu}(170\mu\mathrm{m}) / S_{\nu}(90\mu\mathrm{m})= 1 $ 
is attained at $z \simeq 0.7$
for a pure starburst galaxy, while a cirrus dominated galaxy or a star-forming 
galaxy with a small contribution
from starburst component must be local($z \leq 0.1$).

From Fig.~6 we can also justify the flux calibration based on the IRAS fluxes
of \object{F10507+5723}. If we adopt the ISO 90\,$\mu$m flux determined by PIA and the 
theoretical PSFs, then $S_{\nu}(90\mu\mathrm{m}) / S_{\nu}(60\mu\mathrm{m})= 1.0 $
which is the color of a pure-starburst galaxy, while the ISO flux ratio
$S_{\nu}(170\mu\mathrm{m}) / S_{\nu}(90\mu\mathrm{m})= 1.6 $ is much larger than that
expected for any nearby($z\simeq0$) galaxies. 

The local cirrus dominated galaxies are not likely to be the 
dominant sources
responsible for the fluctuations, since such galaxies are relatively
bright at optical wavelength and hence the source number density in the
Lockman Hole at optical wavelength would exceed the observed one
by more than an order of magnitude: 
as shown in Table~\ref{tab:ncp} we obtained source density of about
$ 2 \times 10^{5}$ sources/sr($60\,\mathrm{sources/deg^{2}}$)
at $S\simeq 100$ mJy. Using the averaged SED of normal galaxies given by
Schmidt et~al. (\cite{Sch97}), the B magnitude of a 100 mJy source at 
90\,$\mu$m is estimated to be about 16 mag. Reported B-band source 
counts above B = 16 mag is only a few sources per square 
degrees(Kirschner et~al. \cite{KOS79}; Ellis \cite{Ell83}).
 Hence the galaxies responsible for the fluctuations must
be heavily extincted at optical wavelengths, which is a well-known 
feature of starburst galaxies.
In conclusion, the major source of the brightness fluctuations is 
most probably the 
star-forming galaxies located at $z < 1$ extincted at optical wavelengths.

In the plot of the 90\,$\mu$m counts in Fig.~\ref{fig:gcounts}, the IRAS counts
are also plotted. These were originally the 60\,$\mu$m counts, in which the flux is
scaled by the factor $S_{\nu}$(90\,$\mu$m)/$S_{\nu}$(60\,$\mu$m)=2.1.
This scaling factor is derived by the IRAS source counts at relatively high
fluxes($ S_{\nu}(100\,\mu\mathrm{m})>1\,\mathrm{Jy}$)
(Rowan-Robinson et~al. \cite{Row86}), and is consistent with the color
of normal galaxies with a partial contribution from starburst. Thus, if such
galaxies dominate the 60\,$\mu$m counts down to 110 mJy, this scaling is valid.
We used the complete differential source counts down to 110mJy by Lonsdale et al. 
(\cite{Lon90}) and Bertin et~al.(\cite{Ber97}), and as shown in 
Fig.~\ref{fig:gcounts} the scaled counts are in good agreement with the model
counts given by Guiderdoni et~al.(\cite{Gui98}).  Since
the integrated counts down to 150 mJy is much larger than the model
counts, the integrated counts must show a steep rise between 150 mJy and 
240 mJy, which, however, could not be observed in the IRAS 60\,$\mu$m counts due
to the source confusion(Hacking and Houck \cite{Hac87}). Further studies from
space like IRIS and SIRTF will judge if this steep rise is real.

\begin{figure}
\vbox{\vspace{0cm}\psfig{figure=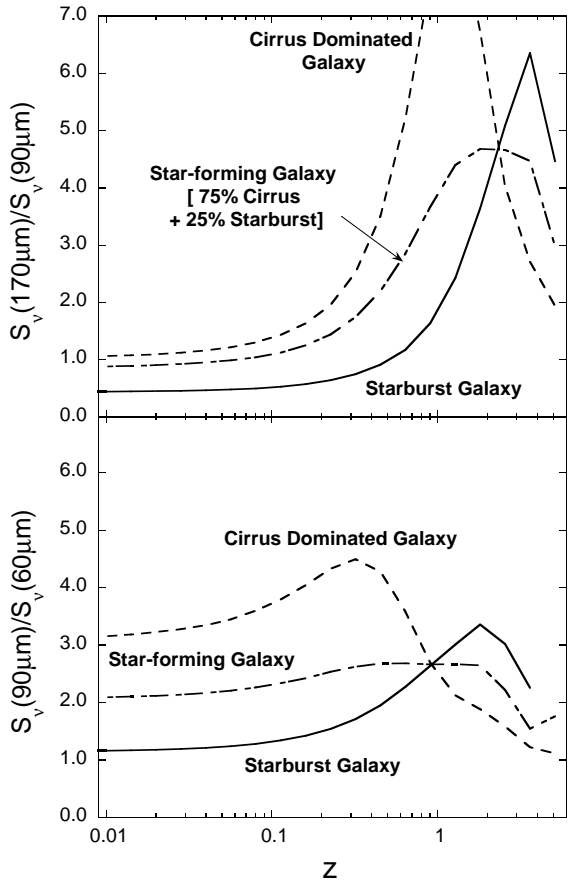,width=8.8cm,clip=}\vspace{0.0cm}}
\parbox[b]{8.8cm}{\caption[]{FIR color vs redshift($z$) relation for
the cirrus dominated galaxy and the starburst galaxy. (Top) the ratio of 
$S_{\nu}(170\,\mu\mathrm{m})$ to $S_{\nu}(90\,\mu\mathrm{m})$, (bottom) 
the ratio of $S_{\nu}(90\,\mu\mathrm{m})$ to $S_{\nu}(60\,\mu\mathrm{m})$.
  The model FIR SEDs are taken from Efstathiou et~al. (\cite{Efs00}) and 
Efstathiou \& Siebenmorgen (\cite{ES00}). The dash-dotted lines
represent the color-$z$ relation of an example of the mixed SED, 
cirrus : starburst = 0.75 : 0.25 at 100\,$\mu$m. }}
 \label{fig:fgcolor}
\end{figure} 

Finally, we mention on the impact of the present work to the cosmic
infrared background(CIB). Hauser et~al.(\cite{Hau98}) reported
an upper limit of 1.1 MJy/sr at 100\,$\mu$m and a positive detection
of the CIB of 1.2 MJy/sr at 140\,$\mu$m and 1.1 MJy/sr at 240\,$\mu$m.
Lagache et~al. (\cite{Lag99a}) reported lower values of the CIB: $0.72\pm0.30$
MJy/sr at 140\,$\mu$m and $0.91\pm0.15$ MJy/sr at 240\,$\mu$m by
further subtracting the dust emission associated with the diffuse
ionized gas.
Lagache et~al.(\cite{Lag00b}) also reported the CIB at 100\,$\mu$m:
$0.78 \pm 0.21$ MJy/sr. 
By summing the fluxes of all detected sources above 150 mJy, we obtained an
integrated brightness of the CIB of 0.031 MJy/sr at 90\,$\mu$m and 0.050 MJy/sr
at 170\,$\mu$m. If we consider the constraints on the number counts
for $S \geq S_{\mathrm{L}}$ shown
in Fig.~\ref{fig:gcounts}, then the integrated brightness would be 0.09 --
0.30 MJy/sr at 90\,$\mu$m for sources above 35 mJy and 0.053 -- 0.15 MJy/sr
at 170\,$\mu$m for source above 60 mJy. Hence, 5 -- 40 per~cent of CIB can
now be attributed to the integrated light of discrete sources above $S_{\mathrm{L}}$
which are responsible for the fluctuations.

\section{Conclusion}

We have studied the characteristics of FIR brightness fluctuations at 90\,$\mu$m
and 170\,$\mu$m in the Lockman Hole, using the power spectrum analysis over the
spatial frequency range of $f=0.05- 1$~arcmin$^{-1}$.
The spectra of PSDs are found to be rather flat at low frequencies,
which differs from that expected for IR cirrus
fluctuations. We interpret the spectral characteristics of PSDs as those of
randomly distributed point sources. The
fluctuations in the Lockman Hole are not dominated by the IR cirrus, and are 
instead most likely due to star-forming galaxies. 
We next showed the constraints on the 
galaxy number counts down to 35 mJy at 90\,$\mu$m and 60 mJy at 170\,$\mu$m,
assuming that only galaxies contribute to the observed fluctuation powers.
This analysis indicates the existence of steep rise in the 
integrated  counts down to these fluxes and especially at 90\,$\mu$m, 
the source density is much 
larger than that expected from the currently available number count models. 

We also found a linear correlation between sky brightnesses at 90\,$\mu$m and 
170\,$\mu$m, even if pixels around the bright sources are masked from the 
images. The slope of the linear
fit is much lower than the color of the IR cirrus, indicating that the
point sources are star-forming galaxies at low or moderate redshift($z<1$). 
These galaxies with fluxes above 35 mJy or 60 mJy also significantly
contribute to the CIB recently reported by Hauser et~al.(\cite{Hau98})
and Lagache et~al.(\cite{Lag00b}).

\begin{acknowledgements}
The authors would like to thank Hiroshi Shibai, Tsutomu Takeuchi, Hiroyuki
Hirashita, and Chris P. Pearson for their extremely useful comments.
HM thanks especially Andreas Efstathiou, who kindly provided us with the 
cirrus galaxy SED before its publication.
\end{acknowledgements}


\end{document}